# GROUND-STATE PROPERTIES OF NEUTRON MAGIC NUCLEI

G. Saxena[1] *, M. Kaushik[2]

**Abstract**--A systematic study of the ground-state properties of the entire chains of even–even neutron magic nuclei represented by isotones of traditional neutron magic numbers [1]$N$ = 8, 20, 40, 50, 82 and 126 has been carried out using relativistic mean-field (rmf) plus Bardeen-Cooper-Schrieffer (BCS) approach. Our present investigation includes deformation, binding energy, two-proton separation energy, single particle energy, rms radii along with proton and neutron density profiles, etc. Several of these results are compared with the results calculated using nonrelativistic approach (Skyrme–Hartree–Fock method) along with available experimental data and indeed they are found with excellent agreement. In addition, the possible locations of the proton and neutron drip-lines, the ($Z$, $N$) values for the new shell closures, disappearance of traditional shell closures as suggested by the detailed analyzes of results are also discussed in detail.

## 1. INTRODUCTION

Several decades ago only about 300 stable or long lived nuclei have been available for the investigation but now with advanced experimental facilities [1], it is possible to examine about 6000 nuclei with comparatively long lifetime. There are various regions of periodic chart which have been found with new magicity or showing disappearance of traditional magicity through different theoretical approaches and confirmed from many experiments. For examples, appearance of the new magic number $N$ = 16 [2–5] along with recent observation for appearance of shell closure at $N$ = 32, 34 have been reported [6–9]. Consequently, drip-line nucleus $^{24}$O and $^{52,54}$Ca are established as new doubly magic nuclei. However, the disappearance of conventional magic numbers has been demonstrated through many communications [10–16].

These findings of new neutron magic numbers have put additional impetus to study already known (conventional) neutron magic number $N$ = 8, 20, 28, 40, 50, 82, 126 and their credibility upto drip-line. In this regard, in year 2015, analyses of $2p$ and $1f$ neutron spin-orbit splitting has been carried out in the $N$ = 20 isotones [17], and $N$ = 40 isotones have been studied by Wang et al. [18]. In addition, In-beam γ-ray spectroscopy of $^{58,60}$Ti has been used to determine the structure of the potentially doubly-magic nucleus $^{60}$Ca with $N$ = 40 [19]. More recently, $β$-decay half-lives for few isotopes of Au, Hg, Tl, Pb and Bi are measured in the mass region around $N$ = 126 [20].

Towards theoretical side, proton and neutron magicity have been studied extensively within the nuclear mean-field models which are widely used in variety of calculations and are based on (i) the Skyrme zero range interaction initially employed by Vautherin and Veneroni [21], and Vautherin and Brink [22], (ii) the Gogny force [23] with finite range, and (iii) the relativistic mean-field model

1) Department of Physics, Govt. Women Engineering College, Ajmer, India.
2) Department of Physics, Shankara Institute of Technology, Jaipur, India.
*Email: gauravphy@gmail.com

formulated by Walecka [24–26], and Boguta and Bodmer [27]. The nonrelativistic models for the low energy nuclear structure physics have been remarkably successful [21–23]. This is essentially due to the fact that the depth of the nuclear potential is below 100 MeV and much smaller than the mass of the nucleon which is nearly 1 GeV. Towards self-consistent theory, generalized method of the energy density functional due to Fayans et al. [28, 29] was proved to be most successful version of the self-consistent theory of finite Fermi systems. In addition, another successful version is the Skyrme–Hartree–Fock (SHF) method with the Hartree-Fock Bogoliubov (HFB)-17 functional given by Goriely et al. [30]. Recently, Fayans functional was used to carry out calculations for spherical nuclei [31] and for deformed nuclei U and Pb for their whole isotopic chain [32]. Moreover, another new developed approach, known as the Barcelona–Catania–Paris (BCP) energy density functional (EDF) [33] and latterly known as the Barcelona–Catania–Paris–Madrid (BCPM) EDF [34] has been developed [35,36] and using it masses of 579 nuclei along with charge radius have been described with a very good accuracy [37].

However, theoretical treatment of magicity has been also described using mean-field theories [38–40] along with their relativistic counterparts [26, 41–45]. Effect of continuum on the pairing energy contribution has been employed by Grasso et al. [39] and Sandulescu et al. [40] within the Hartree-Fock+ Bardeen-Cooper-Schrieffer +Resonant continuum approach. Similarly, the effect of inclusion of positive energy resonant states on the pairing correlations has been examined in detail by Yadav et al. [45]. In recent past, the relativistic mean-field approach has successfully investigated two-proton radioactivity [46], weakly bound drip-line nuclei [47] and magicity [48]. More recently, relativistic mean-field study has been extensively used to describe actinides and superheavy nuclei within covariant density functional theory [49], to calculate decay rates of various proton emitters [50], to study bubble structure [51], to analyze effects of particle-number fluctuation degree of freedom on symmetric and asymmetric spontaneous fission [52] and to calculate neutron capture cross-sections in nuclei near the $N = 82$ shell closure [53].

Encouraged with the latest results of emergence of new neutron shell closure [2–9], and recent study of neutron magic isotones [17–20], in the present investigation we have employed relativistic mean-field plus BCS (RMF+BCS) approach [39–56] to carry out a systematic study of ground-state properties of the entire chains of even-even neutron magic nuclei represented by isotones of traditional (conventional) neutron magic numbers $N = 8, 20, 40, 50, 82$ and 126. One of the main aims of these investigations has been to see if the nuclei belonging to the isotonic chains with neutron numbers $N = 8, 20, 50, 82$ and 126 as well as those of neutron number $N = 40$, whose neutron magicity is traditionally well known in the region close to the line of beta stability, continue to exhibit neutron shell closures as we move away towards the neutron and proton drip-lines. This enables into the possible new aspects of the ground-state properties exhibited by proton-rich as well as the neutron-rich exotic isotones, especially those in the close vicinity of the proton, and neutron drip-lines, respectively. It is worthy and fair to mention here that these neutron magic nuclei have been studied with the use of deformed relativistic mean field (RMF) theory by Geng et al. [57] where pairing correlations have been treated by a simple BCS method with a zero-range δ-force. In addition, RMF theory including deformation degree of freedom and spherical treatment has been employed in Ref [58] for the study of structure of even-even nuclei covering the whole periodic region up to the drip-lines with $8 \leq Z \leq 82$ and $8 \leq N \leq 126$ to identify new proton and neutron magic numbers and doubly magic nuclei for various values of isospins. In this paper, we describe ground-state properties viz. two neutron separation energy, two proton separation energy, radii, density distribution, pairing energy, single particle energy etc. of known neutron magic nuclei with $N =$ 8, 20, 40, 50, 82 and 126 with the use of relativistic mean-field plus BCS approach considering mainly spherical symmetry shapes of nuclei. Furthermore, we will also compare our results with deformed RMF approach and nonrelativistic approach (Skyrme–Hartree–Fock method) [30] to establish general validity of RMF approach. In addition, in order to check the validity of our description for different RMF force parameterizations, we have carried out the spherical RMF+BCS calculations using TMA and NL-SH Lagrangian density which has been equally popular for the relativistic mean-field calculations.

## 2. RELATIVISTIC MEAN-FIELD MODEL

Our RMF calculations have been carried out using the model Lagrangian density with nonlinear terms both for the σ and ω mesons,

$$\begin{aligned}
\mathcal{L} = &\ \bar{\psi}[i\gamma^\mu \partial_\mu - M]\psi \\
&+ \frac{1}{2}\partial_\mu \sigma \partial^\mu \sigma - \frac{1}{2} m_\sigma^2 \sigma^2 - \frac{1}{3} g_2 \sigma^3 - \frac{1}{4} g_3 \sigma^4 - g_\sigma \bar{\psi}\psi \\
&- \frac{1}{4} H_{\mu\nu} H^{\mu\nu} + \frac{1}{2} m_\omega^2 \omega_\mu \omega^\mu + \frac{1}{4} c_3 (\omega_\mu \omega^\mu)^2 - g_\omega \bar{\psi}\gamma^\mu \psi \omega_\mu \\
&- \frac{1}{4} G^a_{\mu\nu} G^{a\mu\nu} + \frac{1}{2} m_\rho^2 \rho^a_\mu \rho^{a\mu} - g_\rho \bar{\psi}\gamma_\mu \tau^a \psi \rho^{\mu a} \\
&- \frac{1}{4} F_{\mu\nu} F^{\mu\nu} - e\bar{\psi}\gamma_\mu \frac{(1-\tau_3)}{2} A^\mu \psi,
\end{aligned}$$

Here the field tensors, $H$, $G$ and $F$ for the vector fields due to ω, ρ and photon are defined through

$$\begin{aligned}
H_{\mu\nu} &= \partial_\mu \omega_\nu - \partial_\nu \omega_\mu, \\
G^a_{\mu\nu} &= \partial_\mu \rho^a_\nu - \partial_\nu \rho^a_u - 2g_\rho \epsilon^{abc} \rho^b_u \rho^c_u, \\
H_{\mu\nu} &= \partial_\mu A_\nu - \partial_\nu A_\mu,
\end{aligned}$$

Furthermore, the symbols M, $m_\sigma$, $m_\omega$ and $m_\rho$, are the masses of nucleon, and that of the σ, ω and ρ mesons, respectively. The superscript "$a$" labels the isospin degree of freedom and runs from 1 to 3. Similarly, $g_\sigma$, $g_\omega$, $g_\rho$ and $e^2/4\pi = 1/137$ are the coupling constants for the mesons, and the photon, respectively, whereas $\tau^a$ are the Pauli isospin matrices. The RMF calculations have been performed assuming 'no-sea' treatment [55] which amounts to neglecting the effects of the Dirac Sea. Based on the single-particle spectrum calculated by the RMF described above, we perform state dependent BCS calculations [59, 60]. For more details of calculations readers can refer [46] and [47].

The set of parameters appearing in the effective Lagrangian include (i) the masses of the nucleons and the mesons M, $m_\sigma$, $m_\omega$ and $m_\rho$, (ii) the coupling constants of the meson fields to the nucleons $g_\sigma$, $g_\omega$, $g_\rho$ and (iii) the parameters $g_2$ and $g_3$ which describe the nonlinear coupling of the σ mesons among themselves, and the parameter $c_3$ which describes the nonlinear self-coupling of the vector meson ω. These have been obtained in an extensive study which provides a reasonably good description for the ground-state of nuclei and that of nuclear matter properties. This set, termed as TMA parameters, has an A-dependence and covers the light as well as medium heavy nuclei from $^{16}$O to $^{208}$Pb [43, 45, 46]. In addition to the use of TMA parameter we also use NL-SH parameter which contains the nonlinear terms for the σ mesons but without the inclusion of nonlinear potential for the ω mesons [44,60]. These two parameters are given below in Table 1.

**Table 1.** Parameters of the Lagrangians TMA and NL-SH together with the nuclear matter properties obtained with these effective forces.

| Parameters | Units | TMA | NL-SH |
|---|---|---|---|
| M | (MeV) | 938.9 | 939.0 |
| $m_\sigma$ | (MeV) | 519.151 | 526.059 |
| $m_\omega$ | (MeV) | 781.950 | 783.0 |
| $m_\rho$ | (MeV) | 768.100 | 763.0 |
| $g_\sigma$ | | $10.055 + 3.050/A^{0.4}$ | 10.444 |
| $g_\omega$ | | $12.842 + 3.191/A^{0.4}$ | 12.945 |
| $g_\rho$ | | $3.800 + 4.644/A^{0.4}$ | 4.383 |
| $g_2$ | (fm)$^{-1}$ | $-0.328 - 27.879/A^{0.4}$ | -6.9099 |
| $g_3$ | | $38.862 - 184.191/A^{0.4}$ | -15.8337 |
| $c_3$ | | $151.590 - 378.004/A^{0.4}$ | |
| **Nuclear Matter Properties** | | | |
| Saturation Density $\rho_0$ | (fm)$^{-3}$ | 0.147 | 0.146 |
| Bulk binding energy/nucleon $(E/A)_\infty$ | (MeV) | 16.0 | 16.346 |
| Incompressibility K | (MeV) | 318.0 | 355.36 |
| Bulk symmetry energy/nucleon $a_{Sym}$ | (MeV) | 30.68 | 36.10 |
| Effective mass ratio m*/m | | 0.635 | 0.60 |

## 3. RESULTS AND DISCUSSION

Our calculations with axially symmetric deformed shapes [42, 44, 46] show that barring only a few cases of isotones, especially those belonging to the $N = 40$ isotonic chain, most of the neutron magic isotones with neutron number $N = 8, 20, 40, 50, 82$ and 126 continue to be spherical in shape and maintain their neutron shell closures for a large range of proton number $Z$ values covering the proton and neutron drip-lines. The $N = 28$ isotonic chain exhibits the same characteristics, most of which are found in the properties of all the chains with the neutron number $N = 8, 20, 40, 50, 82$ and 126. Consequently, to save space and to cover whole periodic region (from $N = 8$ to $N = 126$), the discussion of $N = 28$ isotonic chain has been omitted in the presentation. In order to save further space, the results of isotones have been collectively analyzed and discussed below.

For the isotones having spherical shapes, which is true for most of the neutron magic isotones being considered in the present investigations, we also employ the RMF+BCS approach with spherical symmetry [47, 48] for the analysis of results in terms of spherical single particle wave functions and spherical single particle energy levels to make the discussion of shell closures and magicity etc. more convenient and transparent. It is significant to mention here that within such a spherical framework, the contributions of neutron and proton single particle states to the density profiles, pairing gaps, total pairing energy, etc. which are also equally important in the study of entire chain of isotones can be demonstrated with clarity. For our calculations TMA parametrization [43, 45, 46] which is shown in Table 1, has been used for both spherical and deformed RMF+BCS approach. Further, in order to check the validity of our description for different RMF force parameterizations, we have also carried out the spherical RMF+BCS calculations using NL-SH [44,60] force parameter which has been equally popular for the relativistic mean-field calculations and are also mentioned in Table 1. At various places we will compare our results of both spherical RMF [47, 48] and Deformed RMF [46] approach with the results of SHF method given by Goriely et al. [30] and with available experimental data [61,62].

## *3.1.* *Pairing Energy*

To elaborate persistence of neutron magicity of $N = 8, 20, 40, 50, 82$ and $126$ with proton shell closure (new or conventional), we have displayed the total pairing energy contribution for the chains of isotones as a function of proton number Z with with $N = 50, 82$ and $126$ isotones in the upper panel and $N = 8, 20, 40$ isotones in lower panel of Fig. 1. Since, $N = 8, 20, 50, 82$ and $126$ all are conventionally neutron magic and, therefore, in general for these nuclei the contribution to total pairing energy is mostly from the proton single particle states. With this statement it can be interpreted that for the neutron magic nuclei if total pairing energy vanishes then value of corresponding Z designates proton shell closure.

As can be seen from lower panel of Fig. 1 that pairing energy vanishes at $Z = 2$, 6, 8 and 14 for N = 8 isotones and consequently proton shell closures occur at $Z = 2$, 6, 8 and 14 indicating doubly magic nuclei $^{10}_{2}He_{8}$, $^{14}_{6}C_{8}$, $^{16}_{8}O_{8}$, and $^{22}_{14}Si_{8}$, respectively. In the same way proton shell closures are observed at $Z = 8, 14, 20, 28$ for $N = 20$ isotones with doubly magic nuclei $^{28}_{8}O_{20}$, $^{34}_{14}Si_{20}$, $^{40}_{20}Ca_{20}$, $^{48}_{28}Ni_{20}$. From upper panel of Fig. 1 proton shell closures are found at $Z = 28, 50$ for $N = 50$ isotones in doubly magic nuclei $^{78}_{28}Ni_{50}$, $^{100}_{50}Sn_{50}$. Moreover, for $N = 82$ and $N = 126$ isotones they appear at $Z = 34, 50, 58$ for $^{116}_{34}Se_{82}$, $^{132}_{50}Sn_{82}$, $^{140}_{58}Ce_{82}$ and $Z = 58$ and $Z = 92$ for $^{184}_{58}Ce_{126}$, $^{218}_{92}U_{126}$ respectively as can be seen from upper panel of Fig. 1. Further, for $N = 40$ isotones some non-zero peaks are also observed at proton shell closure $Z = 28$ and $Z = 40$ as can be seen from lower panel. In these cases, contribution from proton shell closure $Z = 28$ and $Z = 40$ remains zero but non zero contribution to pairing energy comes from neutron single particle states indicating non magic character of $N = 40$ for these value of $Z$. However, it is found by spherical RMF calculations [48] that $N = 40$ continues to remain magic from $Z = 16$ to $Z = 26$ only. It is interesting to note here that these nuclei with $N = 40$ and $Z$ from 16 to 26 are actually found with zero quadrupole deformation in the calculations with deformed RMF [46]. It is important to observe from lower panel that for $N = 40$ isotones, pairing energy vanishes at $Z = 20$ giving indications of both proton and neutron shell closure ($Z = 20$ and $N = 40$) in $^{60}_{20}Ca_{40}$. This nucleus $^{60}Ca$ may be the another important doubly magic nucleus near drip-line of Ca next to doubly magic $^{52,54}Ca$ [6–9] for future experiments.

On the other hand, non-zero pairing energy for $N = 50$ isotones at $Z = 20$ can be seen from upper panel of Fig. 1 resulting disappearance of conventional $N = 50$ which is near drip-line of Ca isotopes. Similarly, $N = 126$ does not remain a magic number towards proton-rich side and gives a finite contribution to the total pairing energy as can be reflected from upper panel with the large value of pairing energy near drip-line. Moreover, it is gratifying to note a small peak in upper panel for $N = 50$ isotones and zero pairing energy peak for $N = 82$ isotones at $Z = 34$. These peaks are supporting a subshell closure at $Z = 34$ which one can expect from the isospin symmetry considerations [16] after recent observation of neutron shell closure at $N = 34$ in $^{54}Ca$ [8]. The existence of proton shell closure at $Z = 32$ and $Z = 34$ is a matter of separate discussion in the context of this paper. It is also indulging to note from Fig. 1 that our calculations with both TMA and NL-SH parameters yields similar results and the conclusions from this section are parameter independent.

### *3.2. Two-Proton Separation Energy*

In this section, the variation of two-proton separation energies $S_{2p}$ is depicted with increasing proton number $Z$ for the chains of isotones with neutron number $N = 8, 20$ and $40$ in Fig. 2, and with neutron number $N = 50, 82$ and $126$ in Fig. 3. Deformed RMF+BCS calculations using the TMA force parametrization (denoted by DRMF in all figures) are compared with the available experimental data [61]. Experimental points contain error bars which are sometimes invisible at chosen scale. For isotones with negligible deformation the spherical RMF+BCS calculations (denoted by RMF in all figures) carried out with the TMA and NL-SH force parameters are similar to each other and to the deformed RMF+BCS results. For comparison results obtained by SHB method [30] are also presented. (SHF results of $N = 8$ isotones are not available). The variation in results for the two-proton separation energies $S_{2p}$ also provides important information with regard to the two-proton drip-line. As the proton number $Z$ is increased for a fixed neutron number $N$ the $S_{2p}$ value decreases until it becomes negative whereby we reach the unbound nucleus or the two-proton drip-line. For $N = 8$ this occurs at $Z = 14$ for both the deformed and spherical RMF calculations using the TMA force parameters. This is in accord with the experimental data [61] as can be seen in Fig. 2. The spherical RMF calculations with the NL-SH parametrization predict the two-proton drip-line to occur at $Z = 16$ which is at variance with the results obtained using the TMA parametrization and with experiment. Thus, although, it is in general true that TMA and NL-SH forces yield similar results, slight differences between the two as seen above in the prediction of two-proton drip-line for $N = 8$ isotones does exist. For the isotonic chain with neutron number $N = 20$ all three calculations, the deformed RMF calculation with TMA force and the spherical RMF calculations using TMA and NL-SH forces and SHF method [30], predict the two-proton drip-line to occur at $Z = 26$. Again, this result is in accord with the experimental data [61] as can be seen in Fig. 2. Similarly, other curves given in Figs. 2 and 3 show that the two-proton drip-line for the isotonic chains with neutron number $N = 40, 50, 82$ and 126 occurs at $Z = 44$, $Z = 50$, $Z = 72$ and $Z = 94$ respectively. It is worth to mention here that results of all these calculations are in good match with the results of SHF method [30] as can be seen from the Figs. 2 and 3.

Due to stronger binding of the closed-shell nuclei, an abrupt decrease in the two-proton separation energy $S_{2p}$ is expected to occur for the isotones with $Z$ values located next to the magic $Z$ values. Such a variation can be viewed from the figures of $S_{2p}$ for different isotonic chains given in Figs. 2 and 3.

A close inspection of the figure for the isotonic chain with neutron number $N = 8$, for example, shows rapid decrease in the $S_{2p}$ value at proton numbers $Z = 6$ and 8. Thus one observes, in agreement with experiments [61], the occurrence of new magic numbers corresponding to $Z = 6$ in addition to the traditional proton magic number at $Z = 8$. A similar observation for the $N = 20$ isotonic chain can be made using the corresponding curve in Fig. 2. Again, new shell closure at $Z = 14$ is clearly seen in addition to the traditional shell closure at $Z = 20$. The proton magic numbers which occur for other isotonic chains can be easily inferred from different curves of Figs. 2 and 3. In general, the results of emergence of new proton magic numbers and disappearance of old proton magic numbers in a given isotone, as obtained in previous section, are well corroborated by the sharp and abrupt variations in the $S_{2p}$ values as has been illustrated here.

Moreover, excepting some cases of isotones which are well deformed, the spherical RMF+BCS calculations with TMA and NL-SH force parametrization yield results which are found to be close to the deformed RMF+BCS calculations. The comparison of the results obtained from these three types of calculations with the measurements for the two-proton separation energy of the nuclei in the isotonic chains with neutron numbers $N = 8, 20, 40, 50, 82$ and 126 shows that these are in fairly good agreement with the available experimental data [61] and with the results of SHF method [30] as depicted in the Figs. 2 and 3.

### *3.3.     Two-Neutron Separation Energy*

Analogous to the case of variation of two-proton separation energy with increasing proton number $Z$, which yields information with regard to the two-proton drip-line and also about the shell structures, a study of two-neutron separation energy as a function of decreasing proton number $Z$ provides information on the neutron-rich nuclei of the isotonic chains, and also about the position of two-neutron drip-line. The two-neutron separation energy $S_{2n}$ for the nuclei constituting the isotonic chain with $N = 8, 20, 40, 50, 82$ and 126 obtained in the deformed RMF+BCS calculations has been depicted as a function of proton number $Z$ in Figs. 4 and 5. These figures also show the available experimental data [61] and the results obtained by SHF method [30] for the two-neutron separation energy for the purpose of comparison. Since the calculation of two-neutron separation energy involves the binding energy of nuclei with $N = 6, 18, 38, 48, 80$ and 124, which are in general not spherical, we have not shown in the figures the spherical RMF+BCS calculations. As the proton number $Z$ is gradually decreased for the fixed neutron number $N$, one reaches for the lowest $Z$ at the limit whereby the nucleus with given neutron number $N$ is the most neutron-rich bound nucleus; and beyond that the nucleus becomes unbound against two-neutron emission.

In this sense the nuclei $^{12}_{4}Be_8$, $^{28}_{8}O_{20}$, $^{56}_{16}S_{40}$, $^{70}_{20}Ca_{50}$, $^{114}_{32}Ge_{82}$ and $^{174}_{48}Cd_{126}$ are predicted to be the most neutron-rich bound nuclei in the isotonic chains for the neutron numbers $N$ = 8, 20, 40, 50, 82 and 126, respectively. Further decrease of the proton number below $Z$ = 4, 8, 16, 20, 34 and 50 for the isotonic chain with $N$ = 8, 20, 40, 50, 82 and 126, respectively, causes the two-neutron separation energy to become negative and the corresponding nucleus does not remain bound anymore. The proton numbers $Z$ given above describe the nuclei which define the two-neutron drip-line for the isotonic chains being discussed here. It is seen from the figures that the calculated results for the $S_{2n}$ values are in fairly good agreement with the available experimental data [61] and with the results of SHF method [30] as can be inferred from Figs. 4 and 5.

In the preceding paragraphs, we have discussed the valuable information which can be obtained from a study of two-proton and two-neutron separation energy as a function of proton number Z for the nuclei belonging to different isotonic chains. Variation of two-neutron separation energy for the isotopes of a given nucleus, which belongs to one of the isotonic chains being investigated here, as a function of neutron number can be also instructive if the nucleus whose isotopes are being studied is chosen judiciously.

With this in view and for the purpose of illustration, we have chosen to study the variation of two-neutron separation energy $S_{2n}$ for the $_{84}Po$ isotopes as a function of neutron number N. Since these calculations involve the binding energy of nuclei with proton and neutron numbers $(Z, N)$ and $(Z, N - 2)$, it is possible to correlate these calculations with that of the alpha-decay energy $Q_\alpha$ for the decay of chain of $_{84}Po$ isotopes, which also involves binding energies of nuclei with proton and neutron numbers $(Z, N)$, $(Z - 2, N - 2)$ and that of the α-particle. It is seen that the results of these calculations nicely demonstrate the persistence of neutron magicity at $N$ = 126.

The results of these calculations carried out within the deformed RMF+BCS approach have been displayed in Fig. 6. It shows for the Po isotopes the variation in the two-neutron separation energy $S_{2n}$, and that in the alpha-decay energy $Q_\alpha$ as a function of increasing neutron number $N$. An abrupt change at $N$ = 126 in the two figures is clearly seen. It signifies the evolution of shell closure in the Po isotopes at the neutron number $N$ = 126. The figure also shows the available experimental data [61] for the two-neutron separation energy $S_{2n}$, and for the alpha decay energy $Q_\alpha$. The deformed RMF+BCS results for the $S_{2n}$ values are seen to be remarkably close to the experimental data. Though the calculated $Q_\alpha$ values are somewhat lower than the experimental values, the trend in its variation with increasing neutron number $N$ is indeed very close to that of the measured data [61].

### 3.4. *Radii and Density Distributions*

The results for the rms radii of neutron and proton distributions obtained with the TMA force parametrization have been displayed in Fig. 7 for the isotonic chains with $N = 8$, 20 and 40, and in Fig. 8 for the chains with neutron numbers $N = 50$, 82 and 126. These two figures also display the available experimental data [62] for the purpose of comparison. We have not displayed the results obtained with the NL-SH force as these are similar to those obtained using the TMA force parameters. As mentioned earlier, it may be stated that in the case of well deformed isotones the spherical RMF+BCS description cannot be considered valid. However, in these cases we have also carried out the spherical calculations in order to see the difference between the deformed and spherical descriptions. On the other hand, the isotones which are well deformed occur mostly in the case of neutron submagic $N = 40$ chain only.

It is observed from Figs. 7 and 8 that with increasing number of protons the rms radius for the proton distribution $r_p$ which is calculated from charge radius ($r_p^2 = r_c^2 - 0.64$) increases gradually to have maximum value for the largest allowed $Z$ for which the nucleus in the isotonic chain is bound. For the isotones with negligible deformation, as expected, the deformed RMF+BCS and the spherical RMF+BCS calculations yield similar results for the radii of proton and neutron distributions. In contrast, some neutron-rich as well as proton-rich nuclei, as mentioned above, are substantially deformed. Such a situation, for example, occurs for the nuclei with $Z$ from 36 to 44 in the $N = 40$ isotonic chain in which case one can infer the difference between the results of different calculations for the rms radii as can be seen from Fig. 7. It is observed that the differences are rather small. This implies that the rms radii for the proton and neutron distributions are not affected strongly by deformation.

The variation of the rms radii for the neutron and proton distributions with increasing proton number $Z$ as seen in Figs. 7 and 8 for different isotonic chains clearly shows the enhanced difference between the neutron and proton radii for the isotones on the neutron-rich side, that is for small $Z$ values. Such a neutron skin formation appears to be a general trend for all the isotonic chains. A sharp increase in the neutron radii of the nuclei with minimum $Z$ values of the $N = 8$ and $N = 50$ isotonic chains can be clearly seen in the lower panels of Figs. 7 and 8. The isotones for which these enhancements are maximum correspond to the neutron halo nuclei $^{10}$He and $^{70}$Ca, respectively. Moreover, the nucleus $^{10}$He is located just after the two-neutron drip-line in the $N = 8$ isotonic chain, whereas the other nucleus $^{70}$Ca is found to be located at the top of the two-neutron drip-line for the $N = 50$ isotonic chain. Such dramatic increases signify a halo formation. A neutron halo formation in the nucleus $^{10}$He is experimentally well established. However, in the case of $^{70}$Ca nucleus similar prediction for the halo formation has not been as yet verified [45].

In contrast to the case of neutron-rich isotones, the occurrence of a large proton skin formation in proton-rich isotones is not found, especially in isotonic chains with neutron number $N \geq 40$ as can be seen in Figs. 7 and 8. This is due to the Coulomb effect which limits the value of proton excess in these isotonic chains. However, for the isotonic chains with neutron number $N = 8$ and 20, difference in proton and neutron radii for the two-proton drip-line nuclei is reasonably large and the proton skin formation, though in small measure, does occur as can be seen from Fig. 7.

For the purpose of precise comparison of radii from experimental data [62] and the results of SHF method [30] we have also displayed the charge radii for all the chain of isotones considered here with $N$ = 8, 20, 40, 50, 82 and 126 in Fig. 9. Experimental values have been shown with the error bars. An excellent agreement between non relativistic approach (SHF) and relativistic approach (DRMF) once again can be seen from Fig. 9 for all the isotonic chains. A fairly good agreement is also seen with the available experimental data of charge radii [62].

Detailed features of the results for radii of neutron and proton distributions can be demonstrated by plotting the radial dependence of the neutron and proton densities employing the spherical RMF+BCS description suitable for nuclei which are not deformed. The radial dependence of proton and neutron density distributions for the two-neutron drip-line nuclei, and that of two-proton drip-line nuclei exhibit the spatial characteristics of the densities, respectively, for the neutron-rich and proton-rich nuclei of a given isotonic chain. For the purpose of illustration, we have shown in the upper panel of Fig. 10 the results for the neutron-rich nuclei $^{34}_{14}Si_{20}$ and $^{70}_{20}Ca_{50}$ belonging to the isotonic chains of neutron number N = 20 and N = 50, respectively. These results have been obtained, as stated above, in the spherical RMF+BCS calculations with the TMA force. In both isotones the neutron density is seen to be much more spread beyond the spatial confines of the proton densities. However, in the case of nucleus $^{70}_{20}Ca_{50}$, the neutron density distribution becomes too wide indicating thereby the possible formation of halo structure with loosely bound neutrons. The details of such halo formation has been described in the study of Ca isotopes [45]. We have also shown neutron and proton density distribution of $^{60}_{20}Ca_{40}$. Neutron skin formation can be seen from the curves of $^{60}_{20}Ca_{40}$ and a sharp fall in neutron density further supports magic character of N = 40.

Similarly, in the lower panel of Fig. 10 the results for the proton drip-line nuclei $^{48}_{28}Ni_{20}$ and $^{100}_{50}Sn_{50}$ also belonging to the isotonic chains of neutron number $N$ = 20 and $N$ = 50, respectively, have been displayed. For the proton drip-line nucleus $^{48}_{28}Ni_{20}$ it is seen that the proton density is spread much beyond the neutron core characterizing the situation of a proton-rich nucleus. However, in general it is found that this spatial spread of proton density does not grow enough leading to the formation of the proton halo in nuclei due to the disruptive Coulomb interaction amongst protons. This implies that nuclei in a specific isotonic chain become unbound against two-proton emission with only a small number of addition of protons beyond the line of stability. Thus we do not have the possibility of having bound nuclei which are highly proton-rich.

In contrast, it is possible to have addition of a substantial number of neutrons beyond the line of stability leading to highly neutron-rich nuclei. In general, the proton drip-line does not extend too far as the neutron drip-line does. Consequently, in the case of isotonic chain with neutron number $N$ = 50, the isotones having $Z$ = 52 become unbound against the emission of two-protons. For the $N$ = 50 isotonic chain the proton density of the proton drip-line nucleus $^{100}_{50}Sn_{50}$ shown in the lower panel of Fig. 10 is not spread too far beyond the neutron density boundaries.

In the case of isotonic chains with neutron number $N$ = 82 and $N$ = 126, the proton drip-line nuclei have proton number $Z$ = 72 and $Z$ = 94. These proton drip-line nuclei have rather much less number of protons than the neutrons, and a situation in density profile parallel to the neutron-rich nuclei as seen in the upper panel of Fig. 10 does not occur in the proton drip-line nuclei. Thus we have not presented similar graphs of heavier neutron isotonic chain with $N = 82$ and $N = 126$. In the $N = 40$ isotonic chain, as mentioned before, the nuclei located near the proton drip-line are found to be deformed and thus we have not plotted such a density profile as the spherical RMF+BCS description cannot be considered appropriate for such a deformed case.

### *3.5. Proton Single Particle Energy and Shell Closures*

We have carried out a detailed study of variation of proton single particle energy levels and their variation with addition of protons for all the isotonic chains with $N = 8, 20, 40, 50, 82$ and $126$. To get better insight, we have shown in Fig. 11 such a variation of the proton single particle energies obtained in the spherical RMF+BCS calculation for the $N = 50$ and $N = 82$ isotonic chains. From upper panel of Fig. 11 an appreciable gap between the proton single particle states $2s_{1/2}$ and $1f_{7/2}$ indicates the occurrence of proton shell closure for proton number $Z = 20$. The other gaps indicate the traditional magicity or shell closures at the proton number $Z = 28$ and $50$ as mentioned in the figure.

In a similar way, gaps for traditional proton shell closures are clearly seen in the plot of $N = 82$ isotones in lower panel. It is worth to mention here that a substantial gap is found between proton single particle level $1f_{5/2}$ and $2p_{3/2}$ as can be seen in the lower panel of Fig. 11. This gap is responsible for development of new proton shell closure at $Z = 34$ as has been marked in the lower panel and as in accord with the outcome of section 3.1 for pairing energy. Once again stating that this new shell closure at $Z = 34$ may motivate future experiments after recent observation of neutron shell closure $N = 34$ in $^{54}$Ca [8]. The proton shell closures which are found to occur in other isotonic chains can be obtained by a close inspection of the relevant single particle spectrum. We find that apart from the traditional proton magic numbers, new proton shell closures occur. Thus $Z = 6, 14, 34, 58$ and $92$ corresponds to a new proton shell closure as per proton single particle energy for various other isotones. On the other side, traditional magic numbers disappear due to weakening of neutron single particle gaps responsible for neutron magicity (not shown here). As an example, for $N = 50$, toward neutron-rich side the gap of neutron $1g_{9/2}$ state with next state is found to be only 0.42 MeV at $Z = 20$ and gives rise to disappearance of neutron magicity of $N = 50$ near drip-line. Similarly, $N = 126$ disappears towards proton drip-line with the gap around 2 MeV only.

## 4. SUMMARY

To conclude, the present studies of the nuclei of $N = 8, 20, 40, 50, 82$ and $126$ isotonic chains, carried out within the framework of RMF theory have yielded valuable results related to the ground-state properties, such as binding energy, rms proton and neutron radii, pairing energy, two-proton and two-neutron separation energies, proton and neutron drip-lines and shell closures etc. These results are found to be in reasonably good agreement with the available experimental data [61] and [62] and the results obtained with SHF method given by Goriely et al. [30]. The calculated results for the proton and neutron density profiles, along with the single particle spectra provide further information as regards to the magicity, proton and neutron drip-lines of the isotones as stated above. These results are once again in accord with the available measurements. Thus, the RMF+BCS approach as employed here provides an excellent framework which can be used for extensive studies of nuclei in an economical and fast manner throughout the periodic table as has been successfully demonstrated here for the isotonic chains of the neutron magic nuclei.

However, except light nuclei experimental data for neutron-rich nuclei are scarce in view of present limitations on the radioactive ion beams. Therefore, the neutron drip-lines for most of the chains of nuclei studied here are as yet to be experimentally studied. Study of N = 40 isotones may be crucial for the drip line of Ca isotopes and our identified doubly magic nucleus $^{60}$Ca may be potential candidate for future experiments after $^{52,54}$Ca. Moreover, our finding of proton shell closure Z = 34 may even lead to credibility of isospin symmetry and indeed Z = 34 shell closure may be a testing bench for future experiments.

# AKNOWLEDGEMENTS

Authors would like to thank Prof. H. L. Yadav, Banaras Hindu University, Varanasi, INDIA for his kind guidance and continuous support. The authors are indebted to Dr. L. S. Geng, RCNP, Osaka, Japan for valuable correspondence. One of the authors (G. Saxena) gratefully acknowledges the support provided by Science and Engineering Research Board (DST), India under the young scientist project YSS/2015/000952.

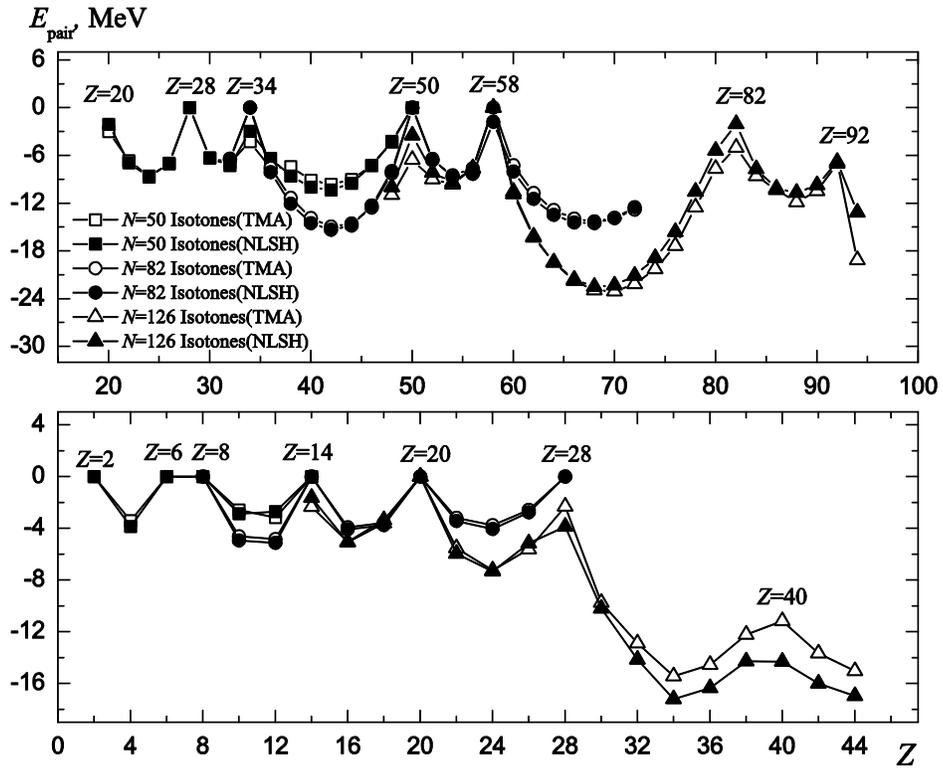

Fig. 1: Upper panel: Present spherical RMF results of pairing energy for the isotones $N = 50$, 82 and 126 obtained with the TMA and NL-SH force parameters. Lower panel: same as upper panel but for $N = 8$, 20 and 40.

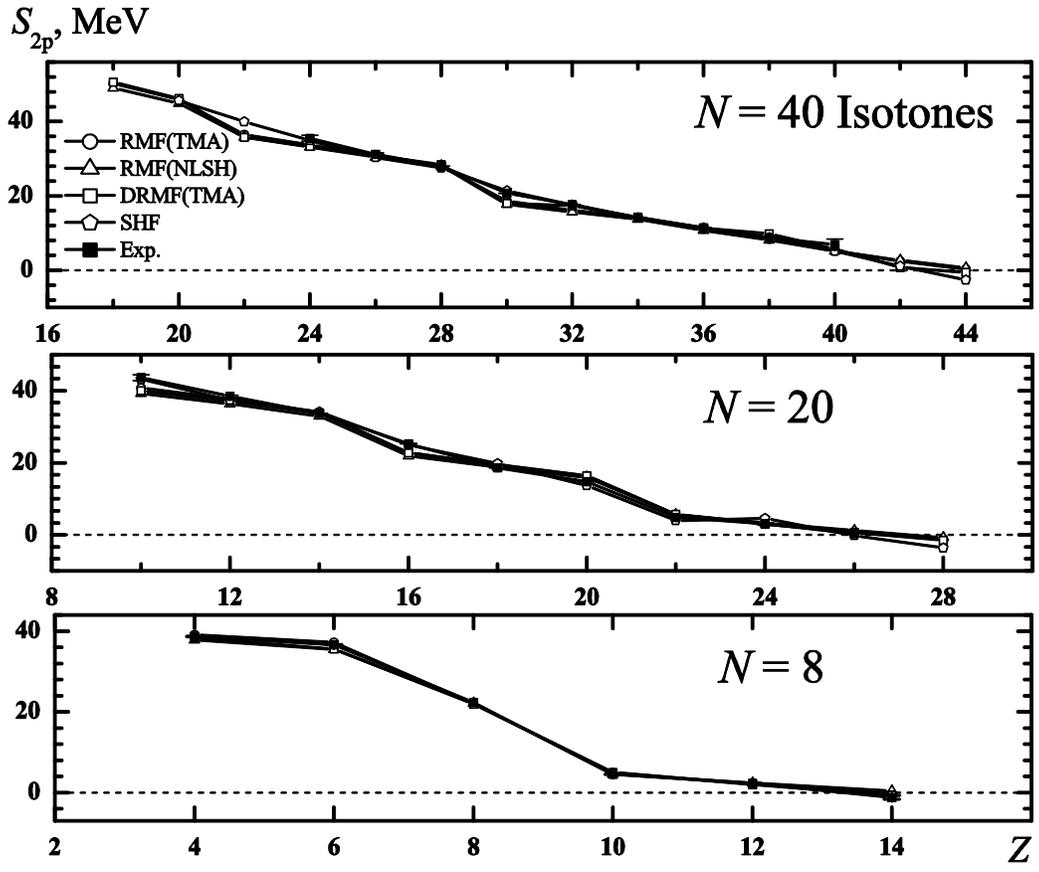

Fig. 2: Variation of the two-proton separation energy $S_{2p}$ for nuclei constituting the $N = 8$, 20 and 40 isotonic chains as a function of increasing proton number Z.

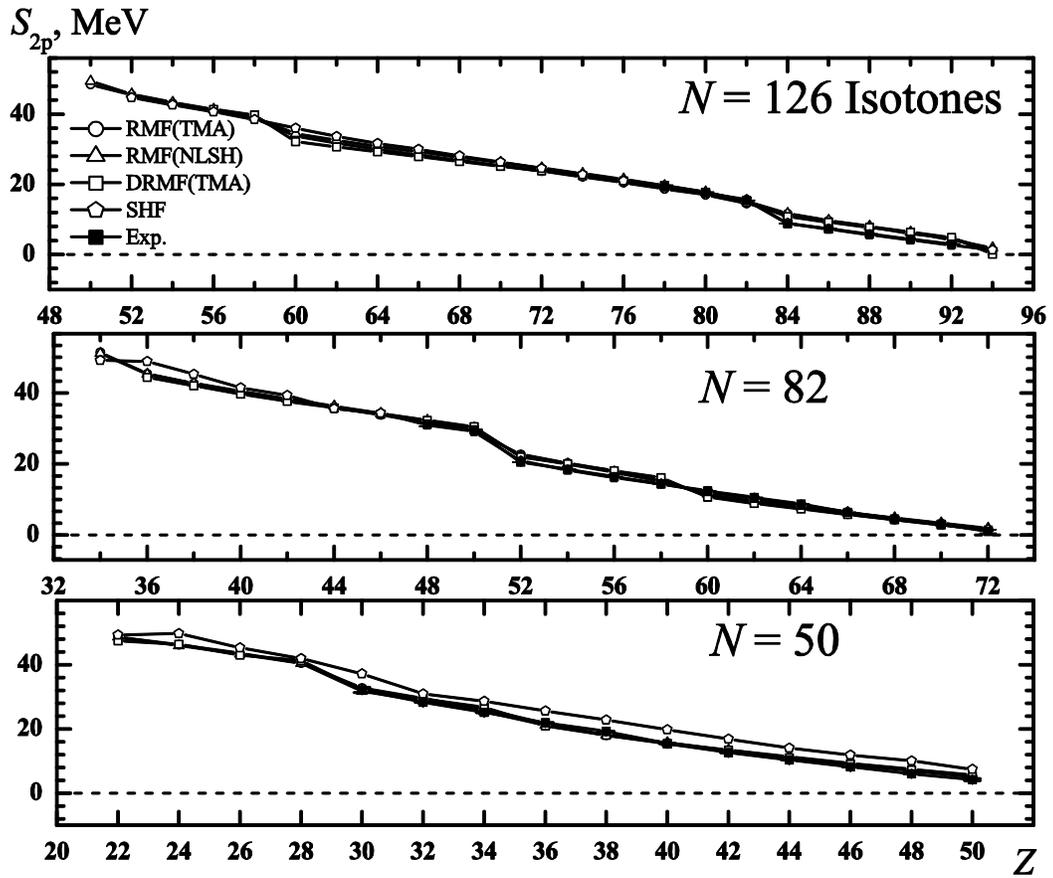

Fig. 3: The same as Fig. 2 but for the isotonic chains with neutron number $N = 50$, 82 and 126.

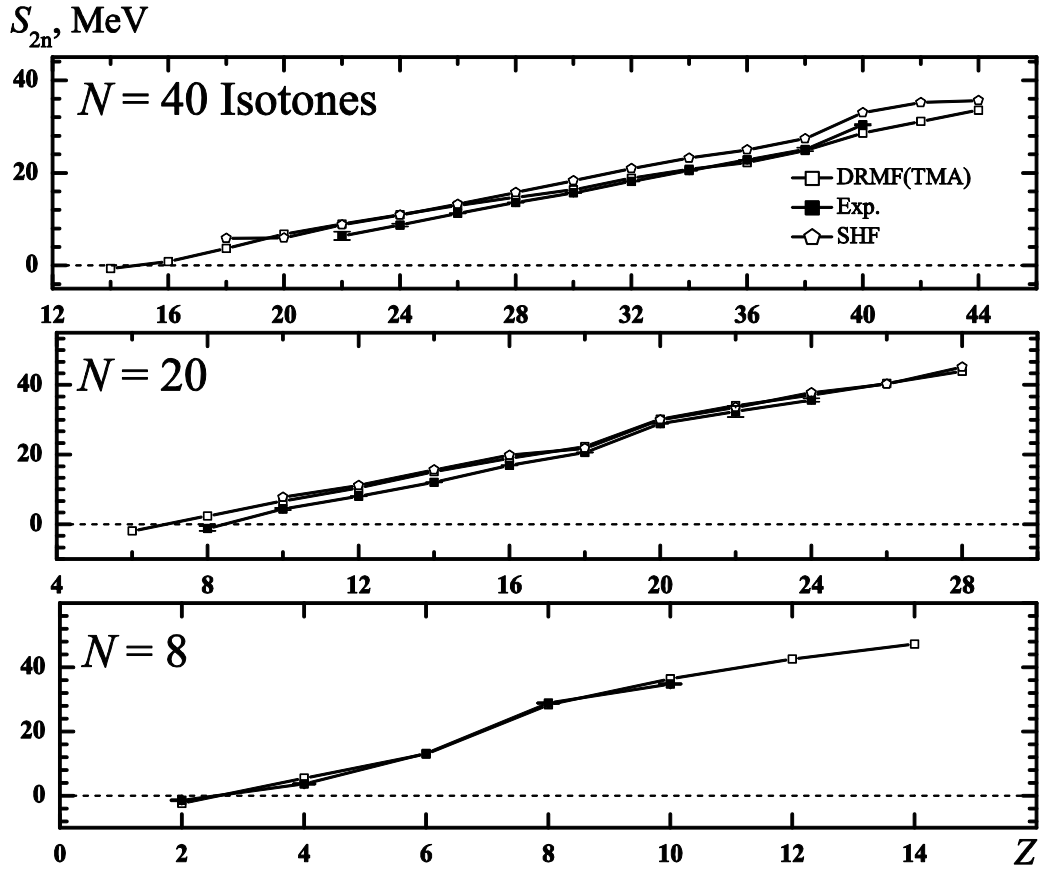

Fig. 4: Variation of two-neutron separation energy $S_{2n}$ for the nuclei constituting the $N = 8$, 20 and 40 isotonic chains as a function of proton number Z.

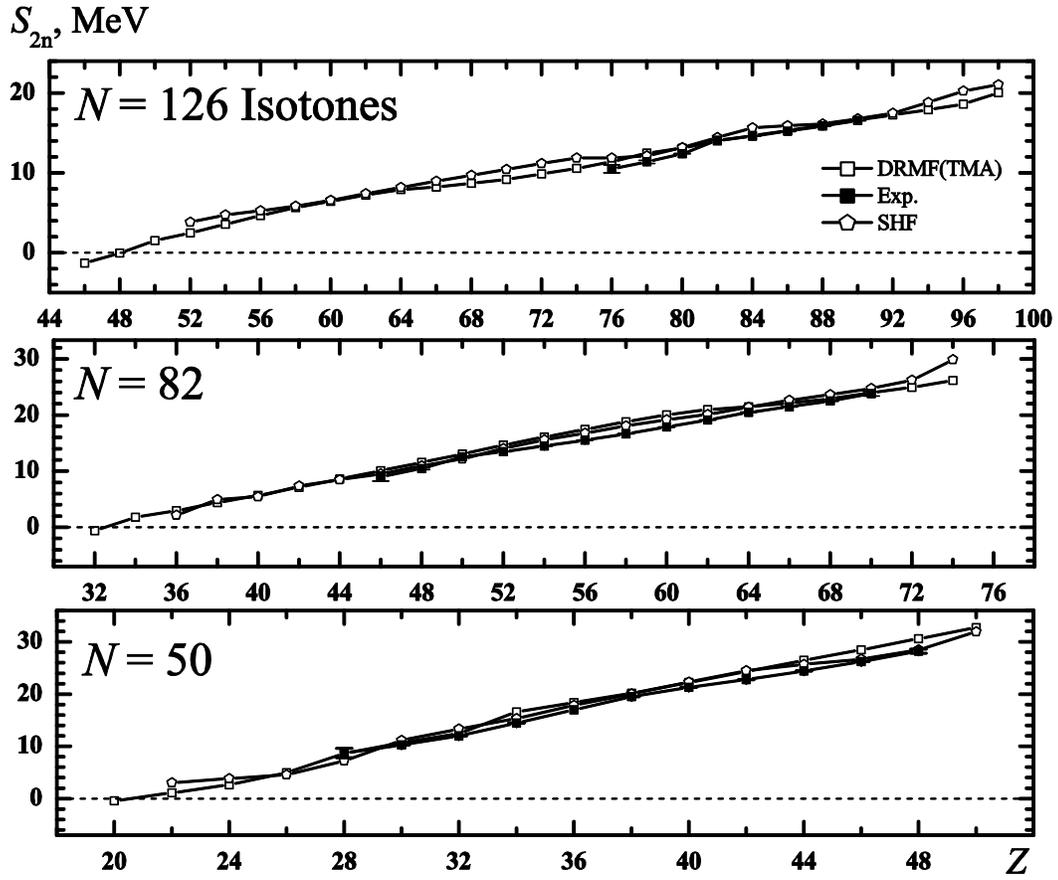

Fig. 5: The same as Fig. 4 but for the isotonic chains with neutron number $N = 50$, 82 and 126.

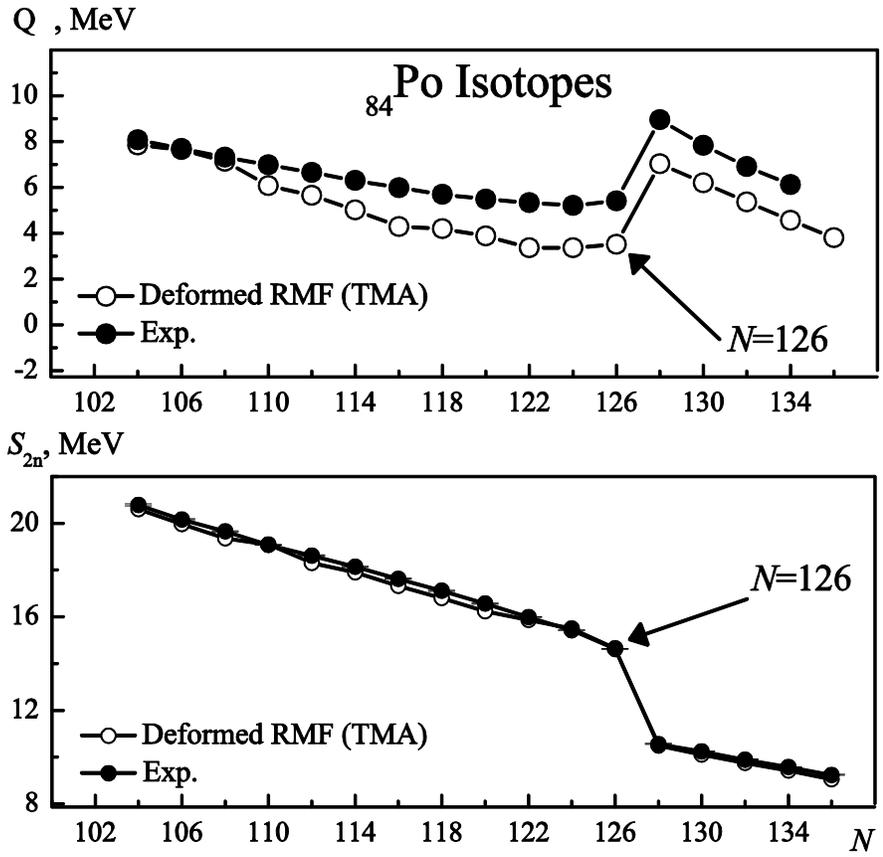

Fig. 6: Alpha decay energy $Q_\alpha$ (upper panel) and results of calculated two-neutron separation energy $S_{2n}$ (lower panel) for the $_{84}$Po isotopes within the deformed RMF+BCS approach are compared with the available experimental data [61] as a function of increasing neutron number N.

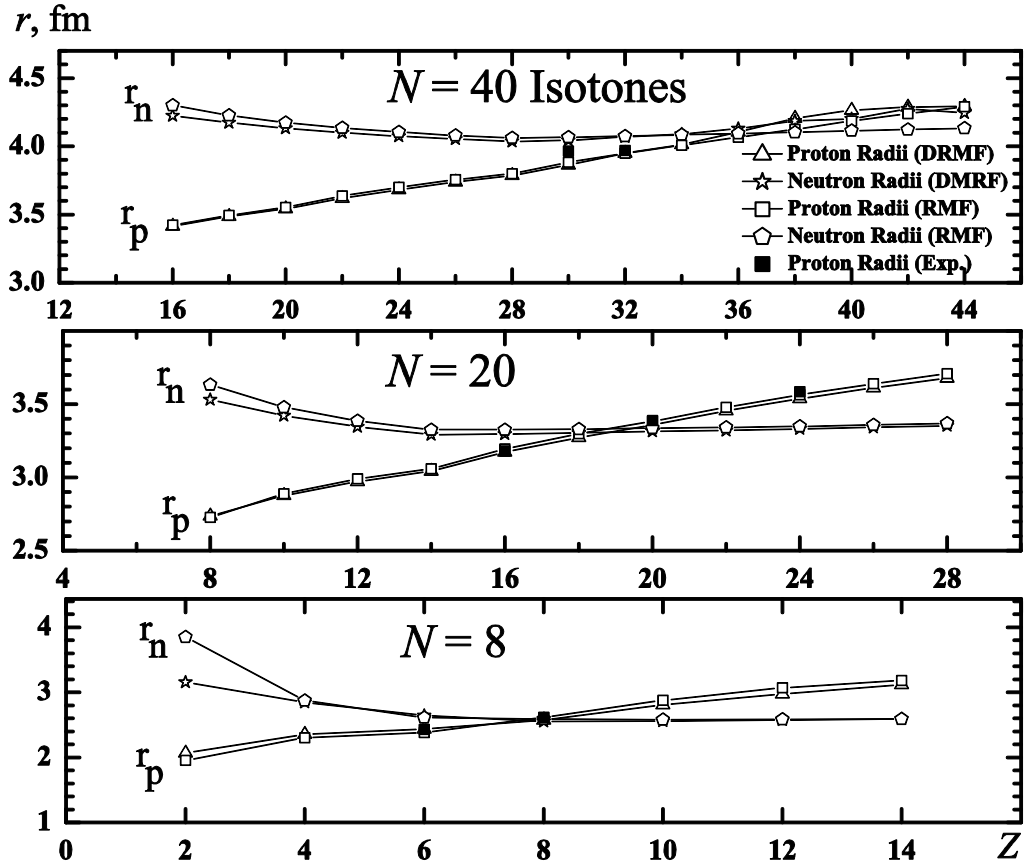

Fig. 7: The rms radii of neutron and proton distributions, $r_n$ and $r_p$, respectively, for the nuclei constituting the $N = 8$, 20 and 40 isotonic chains obtained with the deformed RMF+BCS calculations using the TMA force parameters are compared with the available experimental data. The figure also shows the results of neutron and proton radii obtained in the spherical RMF+BCS approach with the TMA force parameters.

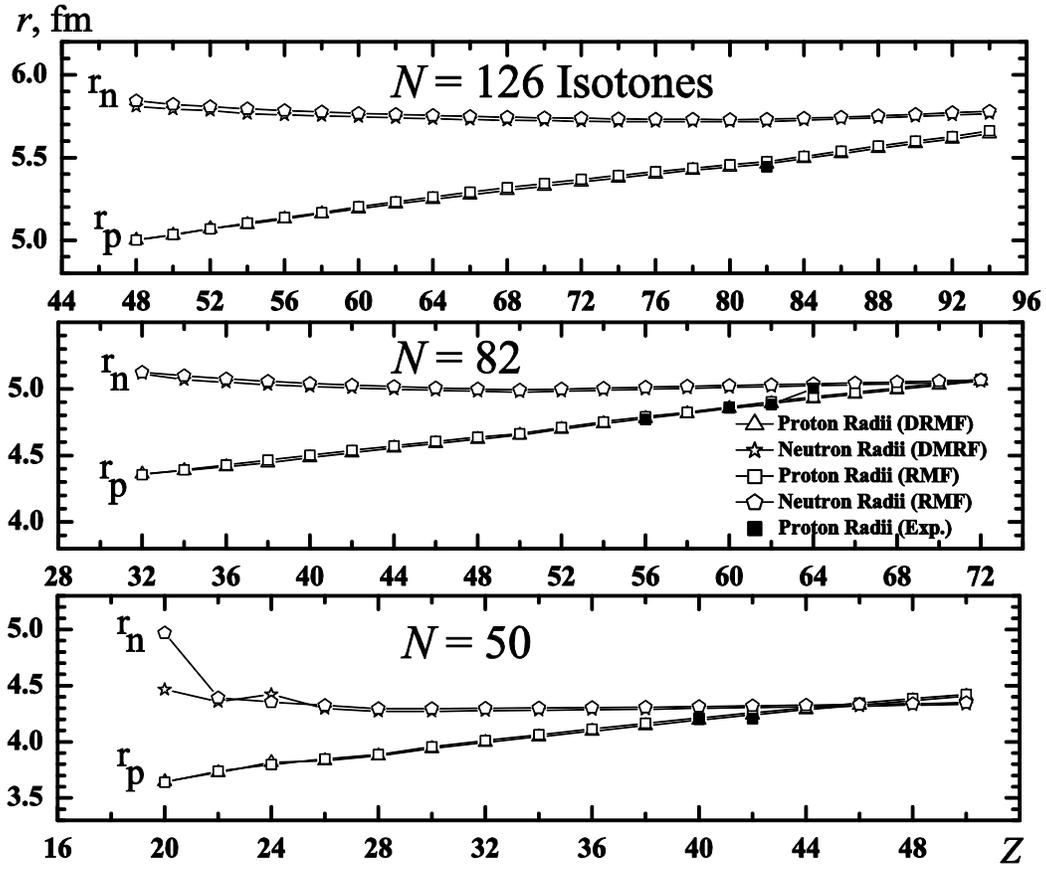

Fig. 8: The same as Fig. 7 but for the isotonic chains with neutron number $N = 50$, 82 and 126.

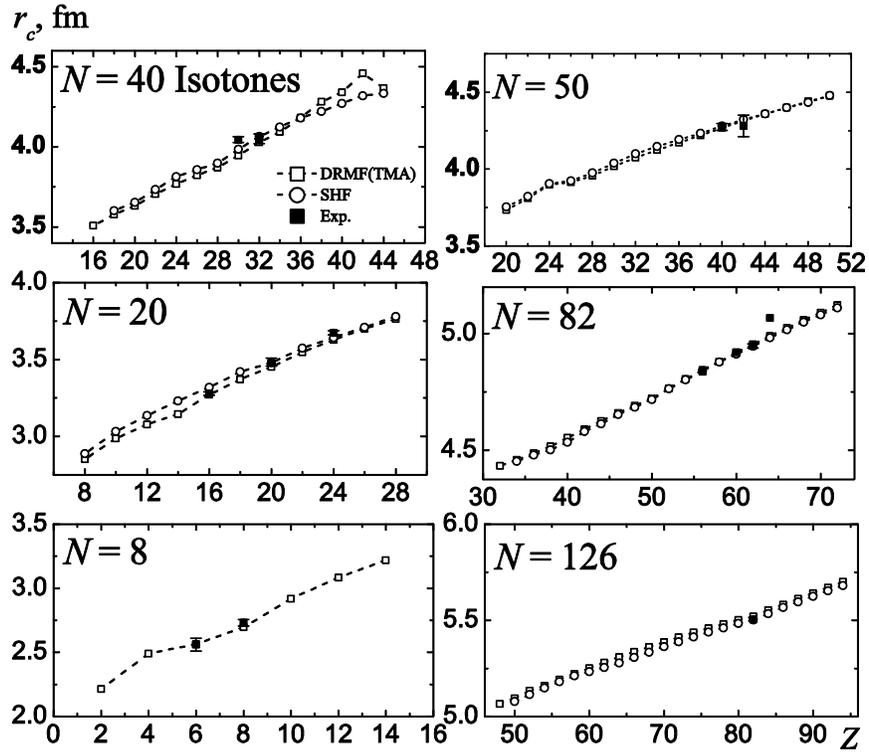

Fig. 9: The charge radii for the nuclei constituting the $N$ = 8, 20, 40, 50, 82 and 126 isotonic chains obtained with the deformed RMF+BCS calculations using the TMA force parameters are compared with the available experimental data. The results of charge radii obtained by SHF [30].

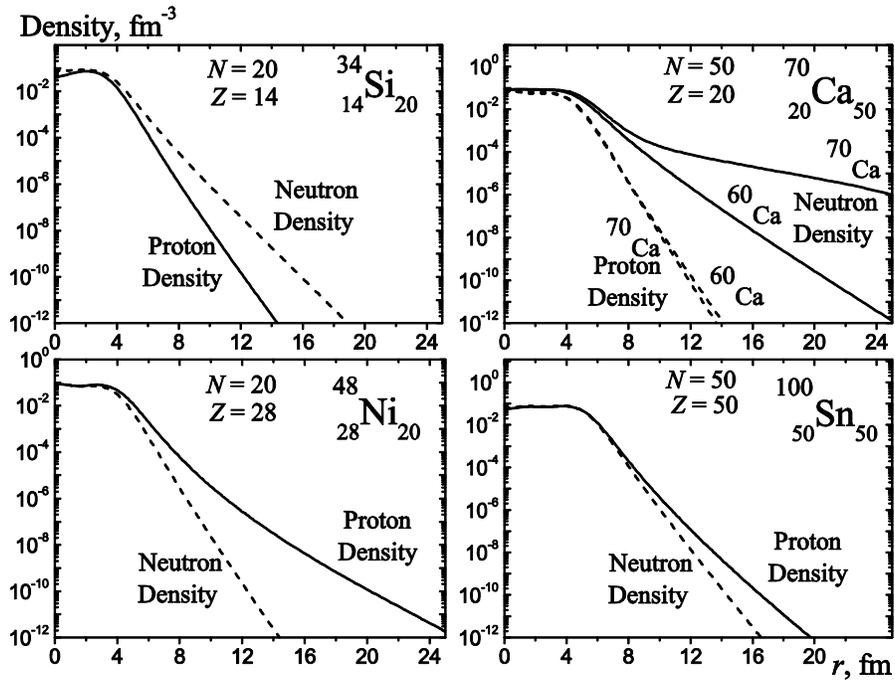

Fig. 10: Results for the proton and neutron density distributions obtained in spherical RMF+BCS calculations with the TMA force for the two cases of N = 20 and N = 50 isotonic chains.

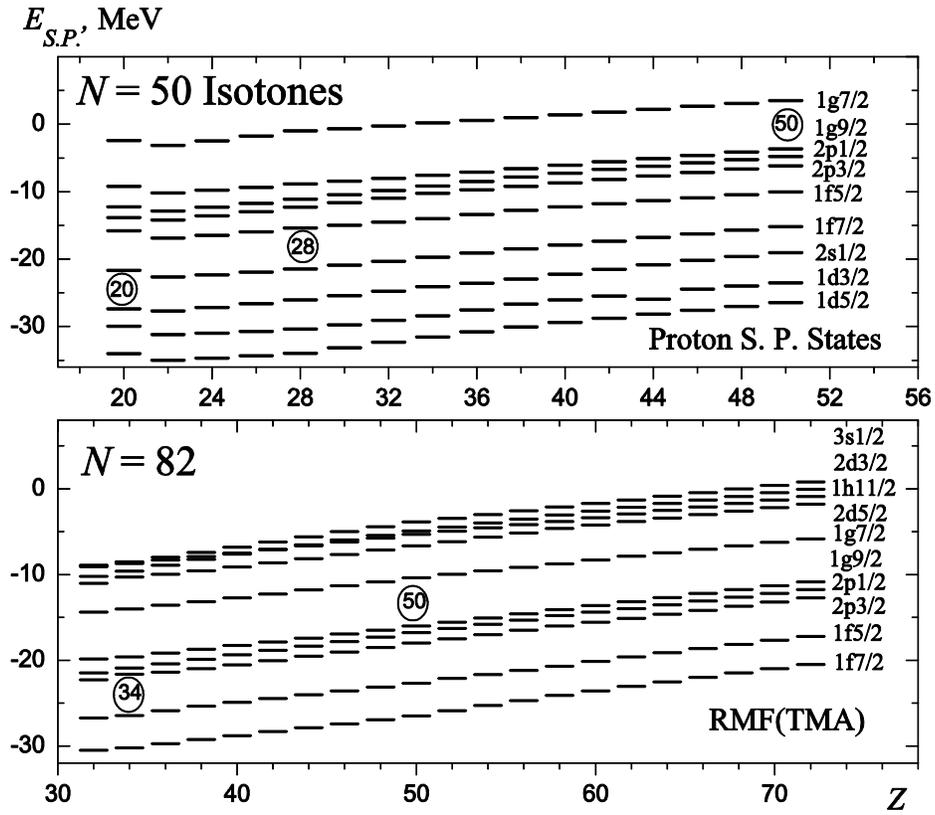

Fig. 11: Proton single particle energies of $N = 50$ and $N = 82$ isotones obtained in the spherical RMF+BCS calculations with the TMA force in the upper and lower panel respectively.